\documentclass[
  journal=pasa,
  manuscript=research-paper, 
  year=202X,
  volume=YY,
]{cup-journal}

\usepackage{amsmath}
\usepackage{amssymb,microtype,siunitx,booktabs}
\usepackage[skip=10pt]{parskip}
\usepackage{relsize}
\usepackage{xcolor}
\usepackage[dvipsnames]{xcolor}

\usepackage{soul}
\soulregister\citeA7
\soulregister\cite7
\soulregister\ref7

\usepackage[finalnew]{trackchanges}
\addeditor{JE}

\title{An Analysis of a Coronal Mass Ejection Leading Edge by Means of Multi-Spacecraft-in-Beam Phase Scintillation}

\author{J. Edwards}
\affiliation{University of Tasmania, Hobart, 7004, TAS, Australia.}
\alsoaffiliation{CSIRO, Space \& Astronomy, Epping, 1710, NSW, Australia.}
\email[J. Edwards]{jasper.edwards@utas.edu.au}

\author{G. Molera Calvés}
\affiliation{University of Tasmania, Hobart, 7004, TAS, Australia.}

\author{J. Morgan}
\affiliation{CSIRO, Space \& Astronomy, Bentley, 6102, WA, Australia.}

\author{M. Cheung}
\affiliation{CSIRO, Space \& Astronomy, Epping, 1710, NSW, Australia.}

\received {dd Mmm YYYY}
\revised  {dd Mmm YYYY}
\accepted {dd Mmm YYYY}
\published{XX XX 202X}


\begin{document}

\begin{abstract}
A Coronal Mass Ejection (CME) was detected crossing the radio signals transmitted by the Mars Express (MEX) and Tianwen-1 (TIW) spacecraft at a solar elongation of $4.4^{o}$. The impact of the CME was clearly identifiable in the spacecraft signal SNR, Doppler noise and phase residuals observed at the University of Tasmania's Very Long Baseline Interferometry (VLBI) antenna in Ceduna, South Australia. The residual phases observed from the spacecraft were highly correlated with each other during the transit of the CME across the radio ray-path despite the spacecraft signals having substantially different Doppler trends. We analyse the auto- and cross-correlations between the spacecraft phase residuals, finding time-lags ranging between 3.18-14.43 seconds depending on whether the imprinted fluctuations were stronger on the uplink or the downlink radio ray-paths. We also examine the temporal evolution of the phase fluctuations to probe the finer structure of the CME and demonstrate that there was a clear difference in the turbulence regime of the CME leading edge and the background solar wind conditions several hours prior to the CME radio occultation. Finally, autocorrelation of the MEX two-way radio Doppler noise data from Ceduna and closed-loop Doppler data from ESA's New Norcia ground station antenna were used to constrain the location of the CME impact along the radio ray-path \add{to a region 0.2 AU from the Sun, at a heliospheric longitude consistent with CME origin at the Sun. The results presented demonstrate the potential of the multi-spacecraft-in-beam technique for studying CME structures in great detail, and providing measurements that complement the capabilities of future solar monitoring instruments.}
\end{abstract}

\section{Introduction}
Since the first experiments measuring plasma induced spectral broadening of the radio telemetry signal from the Pioneer 6 interplanetary probe \citep{goldstein1969superior}, spacecraft radio propagation techniques have been used extensively to probe both the solar corona and the heliosphere. \add{A comprehensive overview of the different techniques and early spacecraft radio propagation experiments can be found in }\citet{bird1982coronal}. More recent reviews on the specific techniques and advances since the early experiments can be found in \citet{Remus2014} (phase scintillation, Doppler and ranging), \citep{efimov2008radial} (amplitude scintillation) and \citet{kooi2022modern} (Faraday rotation). To date, these techniques still offer a highly effective way of probing the inner heliosphere and structures within the solar corona, especially very close to the Sun, where in situ spacecraft cannot ordinarily or routinely obtain measurements. Results from these types of experiments are pertinent to both fundamental plasma physics and to operational space weather science and forecasting.

A variation of the radio propagation experiments that is of particular interest is the ability to measure bulk plasma structure and velocity by observing the delay in scintillation effects, including fluctuations in amplitude, frequency and phase, between multiple adjacent radio paths \citep{efimov1981observations}. Experiments of this delineation have typically been conducted with only two ground station receivers, such as the United States National Aeronautics and Space Administration (NASA) Deep Space Network (DSN) \citep{armstrong1979interplanetary, bird2002outer}. However, more recently, experiments employing multiple Very Long Baseline Interferometry (VLBI) antennas have also been used, providing many radio ray-paths through the interplanetary plasma \citep{ma2021vlbi, Ma2022}. The number of ray-paths combined with the high temporal resolution of VLBI antenna data means the plasma structure and variability can be examined in great detail, an ability which is not possible with only one antenna.

Multiple radio ray-paths can also be achieved by having two or more spacecraft signals arrive at a single receiver. This variation of the technique was first planned for the dual-spacecraft International Solar Polar Mission (ISPM) which would have featured a continuous two spacecraft solar occultation that could have been tracked at a single DSN station with both spacecraft in-beam \citep{ bird1982coronal}. The ISPM was subsequently abandoned and then later evolved into the single-spacecraft Ulysses mission, which retained a productive radio propagation experiment \citep{bird1994coronal}, albeit without testing the dual-spacecraft method. Since the abandonment of the ISPM, instances of the multi-spacecraft technique have been implemented. For example, both configurations (multi-spacecraft and multi-ground station) were exploited during the operational lifespan of Venera-15 and -16 (orbiting Venus) by \citet{yakovlev1989occultation} to estimate the bulk velocities and the stability of plasma inhomogeneities as they were transported between the radio ray-paths. More recently, plasma drift across the radio ray-paths of more significantly separated spacecraft with different ground stations -- Akatsuki (orbiting Venus) and BepiColombo (en route to Mercury), was conducted by \citet{cappuccio2024probing}. 

Planets hosting multiple active spacecraft such as Mars, where the European Space Agency (ESA) Mars Express spacecraft (MEX) and \add{China National Space Administration} (CNSA) Tianwen-1 spacecraft (TIW) currently operate, or Jupiter, where NASA's Europa Clipper mission and ESA's Jupiter Icy Moons Explorer (JUICE) mission will be operational simultaneously, present an interesting opportunity to conduct multi-spacecraft-in-beam experiments. The Planetary Radio and Interferometry Experiment (PRIDE) has been conducting radio experiments of the MEX spacecraft since 2013 \citep{Kummamuru2023} and continues to observe the spacecraft signal, particularly targeting superior solar conjunctions \citep[hereafter referred to as Paper I]{edwards2025studying}. PRIDE is also a formal component of the JUICE mission and large amounts of radio science data will be collected with VLBI antennas for precision orbit determination \citep{gurvits2023planetary}. Both the current Mars missions and next generation Jupiter missions offer the possibility to use the multi-spacecraft-in-beam experiments to study space weather, as proposed by \citet{said2025simultaneous}.

In addition to studying the solar corona and background solar wind, radio propagation techniques are useful for probing interplanetary disturbances including Coronal Mass Ejections (CMEs). CMEs are the most significant interplanetary disturbances, comprised of large magnetic field structures that propagate away from the Sun carrying plasma material with them. The modern understanding of the structure of CMEs, the classic three-part model comprised of a leading edge (sometimes developing a shock front), a lower density `cavity' region and a dense highly turbulent main core structure, is largely based on observations of Thompson scattered optical light from the inner heliosphere \citep{howard2011coronal}, observed with coronagraph instruments, \add{detections of low frequency non-thermal radio emissions} \citep{gopalswamy2011coronal,webb2012coronal}, in situ measurements such as those presented in \citet{nieves2013inner} and results from spacecraft radio propagation experiments.

A general model for CME structure developed from spacecraft radio differential frequency and propagation delay techniques was proposed by \citet{patzold2012coronal}. According to this model, the general structure of a CME is comprised of an initial smaller density increase, the leading edge, which precedes a main larger and denser outward moving front. The leading edge acts as a plough, accumulating coronal material into the CME `sheath' structure as it propagates outward through the corona and is typically characterized by either large variations of increasing electron content or by a roughly linear and steady increase, without significant fluctuation. Behind the leading edge, the electron content increases sharply at the CME core region. The leading edge and core structure are also often separated by a region of reduced electron content, or in some cases, variations of electron content, which corresponds to the cavity region seen in coronagraph images. The core structure is typically followed by a series of smaller fronts, separated from each other with turbulent material. Scintillation observations have revealed this region can also have significantly reduced electron content which is gradually replenished by material from the neighbouring corona. Radio observations by \citet{ando2015internal} (scintillation) and \citet{ jensen2018plasma} (Faraday rotation) observed a second density peak in this trailing region, accompanied by an enhancement of the velocity. These were attributed to converging fast plasma ﬂow associated with the magnetic reconnection behind the CME main structure. Scintillation observations of a CME structure by \citet{lynch2002comparison} also revealed an exceptionally dense filament like structure, fine enough that it was not visible in coronagraph imagery, however, whether this structures developed within the CME or whether it was pre-existing and was pushed or transported along with the CME front remained unclear.

\defcitealias{edwards2025studying}{Paper~I}

Observing CMEs using the PRIDE technique was demonstrated at large solar elongation (SEP angle of $17^{o}$) by \citet{Moleracalves2017}. More recently, the PRIDE method was demonstrated as being effective for observing exceptionally close to the Sun ($0.7^{o}$ or $2.6\ R_{\odot}$) in \citetalias{edwards2025studying}, which performed as a proof of concept for observing transient events with density comparable to that of the inner corona. These experiments demonstrated that PRIDE observing method was robust for observing CME structures both very close to the Sun and much further away from the Sun, beyond the field of view of the coronagraph instruments.

Though the general physical driving mechanisms and structure of CMEs are relatively well understood, \remove{however, }there remain open questions regarding the development, morphology and evolution of CMEs, especially as they propagate further from the Sun \citep[and references therein]{temmer2023cme}. From the perspective of space weather forecasting, the key focuses are on the formation of shocks ahead of CME structures that result in enhancement of the CME geo-effectiveness and the influence of drag (induced by the solar wind) and deflection (caused by solar or planetary magnetic fields) which may alter the trajectory of the CME \citep{manoharan2000radial,  manchester2017physical}. Resolving these open questions is of particular interest to current research because CMEs pose a significant hazard to infrastructure on Earth and in space. Developing a better understanding of the physics of these events, the unique cases and instances where CMEs may deviate from the general model is essential for space weather prediction. Radio propagation experiments, including the multi-spacecraft-in-beam technique, provide an effective way of probing CMEs to help answer these questions.

In this paper, we present simultaneous measurements of the phase scintillation of radio signals from the MEX and TIW spacecraft caused by a CME occurring on December $3^{rd}$ during the 2023 Mars solar conjunction. The paper is structured as follows. The relevant theory and equations of phase scintillation as a technique for probing interplanetary plasma and, more specifically, CMEs, is discussed in \S \ref{background}; an overview of the radio science experiment including the data reduction and post processing methods is also presented in this section. A discussion of the radio science results, including context provided by ancillary data from LASCO C3 imagery, analysis of the CME leading-edge including phase residuals, plasma drift delays, velocity and location, total electron content (TEC), and turbulence spectral index is presented in \S \ref{results and analysis}; finally, the concluding remarks are given in \S \ref{conclusion}. 

\section{Method}\label{background}
\subsection{Relevant Phase Scintillation Theory} \label{background_scint}
\add{The phenomenon of interplanetary scintillation generally refers to the fluctuations in amplitude, frequency and phase of a radio signal which are caused by propagation of the radio signal through the interplanetary plasma medium.} Inherent phase changes of an electromagnetic wave propagating through an inhomogeneous plasma medium, such as the interplanetary solar wind, can be fully understood by changes in refractive index, resulting from changes in electron density along the radio propagation path. The phase shift for an individual density irregularity enhancement of characteristic scale size, $a$, is given by:

\begin{equation}
\begin{aligned}\label{densityEq}
\Delta \phi = r_{e} \lambda a \Delta n_{e} \\
\end{aligned}
\end{equation}

where $r_{e}$ is the classical electron radius ($2.82 \times 10^{-15}$m), $\lambda$ is the radio wavelength of the observed signal ($3.6$ cm here\add{, corresponding to 8.4 GHz}) and $\Delta n_{e}$ is the excess electron density relative to the ambient medium \citep*{alurkar1997solar, thompson2017interferometry}. Contributions to the phase fluctuations occur from density fluctuations occurring in a region much larger than the Fresnel scale and depend on the distance between the observing antenna and the scattering plasma \citep{narayan1992physics}. 

The degree of phase scintillation can then be characterised by the phase stability or scintillation index \citep{duev2012spacecraft, MoleraCalves2014}:

\begin{equation}
\begin{aligned}\label{eq:scintillation_index}
\sigma_{Sc}=\Biggr[ \int_{f_{l}}^{f_{h}}S_{\phi}(f)\,df \Biggr]^{1/2}
\end{aligned}
\end{equation}

\add{where $S_{\phi}(f)$ is the power spectrum (the two-sided spectral density) of the residual phase after the Doppler compensation, the instantaneous phase fluctuations $\phi(t)$,} in units of radians squared per Hz. The high frequency limit, $f_{h}$ (Hz), is dictated by the phase scintillation dropping off at higher frequencies and becoming obscured by the system noise (chosen to be 3.003 Hz). The low frequency limit, $f_{l}$ (Hz), is constrained by factors including the (finite) duration of the phase time series, $T$, and the Doppler compensation process, \add{chosen in this study to be 0.003 Hz for data durations of 20-minutes and 0.01 Hz for data durations of 5-minutes. Both limits, $f_h$ and $f_l$, have been chosen to produce measurements consistent with previous studies}. The evolution of $\sigma_{Sc}$, determined for a 5-minute ($T=5$) time series, can also be used to describe the structure of the plasma fluctuations at a higher resolution than when aggregated over the 20-minute duration ($T=20$).

Determination of the degree of scintillation using $S_{\phi}(f)$, instead of parameters like the temporal phase structure function or the phase covariance, is useful because the error in power is uncorrelated between neighbouring frequencies \citep{wheelon2001electromagnetic}. This allows clear identification and filtering of the system noise. An analysis of the noise impacts of different parameters, including aperture diameter, atmospheric effects and antenna system phase noise, specific to the range of VLBI antennas used in the PRIDE space weather experiments was done by \citet{kummamuru2025estimating}. \add{The inherent system noise introduced by instrumental or atmospheric factors, identified by comparisons of the spacecraft signal to injected phase calibration tones, is approximately a $\pm 0.011$ radian contribution to the phase RMS. Close to the Sun, the scintillation effects extend into the noise band (above $f_h$) and the contributions from the ionosphere are effectively negligible, as demonstrated in} \citetalias{edwards2025studying}. \add{We define the uncertainty of $\sigma_{Sc}$ as $\frac{\sigma_{Sc}}{\sqrt{N}}$ where N is the number of samples of the phase residuals (N=10800 for T=5 and N=46800 for T=20). The uncertainty in the gradient of  $S_{\phi}(f)$ is the standard error of the linear regression fit over the range from $f_l$ to $f_h$, which, for the observations presented here, was found to be a contribution of no more than a $\pm 0.01$ to the slope, determined for data segments of 5-minutes ($T=5$).}

The quantity $\sigma_{Sc}$, determined for a 20-minute time series, can then be used to estimate the total electron content (TEC), the electron content integrated along the complete ray-path, of the background solar wind through the relationship empirically derived in \citet{Kummamuru2023}: 

\begin{equation}
\begin{aligned}\label{eq:TEC}
TEC & = K \sigma_{Sc \ (T=20)} \\
K & = 2390 \pm 359 \ \ \ \textnormal{ TECU rad$^{-1}$}
\end{aligned}
\end{equation}

where $K$ is an empirically determined scaling factor, determined from over 1000 observations of MEX, with a magnitude of $2390 \pm 359 \ \ $TECU rad$^{-1}$. \add{The scale of $K$ is primarily dependent on the duration over which the scintillation index has been calculated, the low and high frequency bounds ($f_l$ and $f_h$) applied to $S_{\phi}(f)$ and the distance to the target spacecraft} (see, for example, \citet{MoleraCalves2014} for a similar relationship for scintillation observations of Venus Express spacecraft).

\subsection{Multi-Spacecraft-In-Beam Technique}
\begin{figure}[hbt!]
\centering
\includegraphics[width=0.9\linewidth]{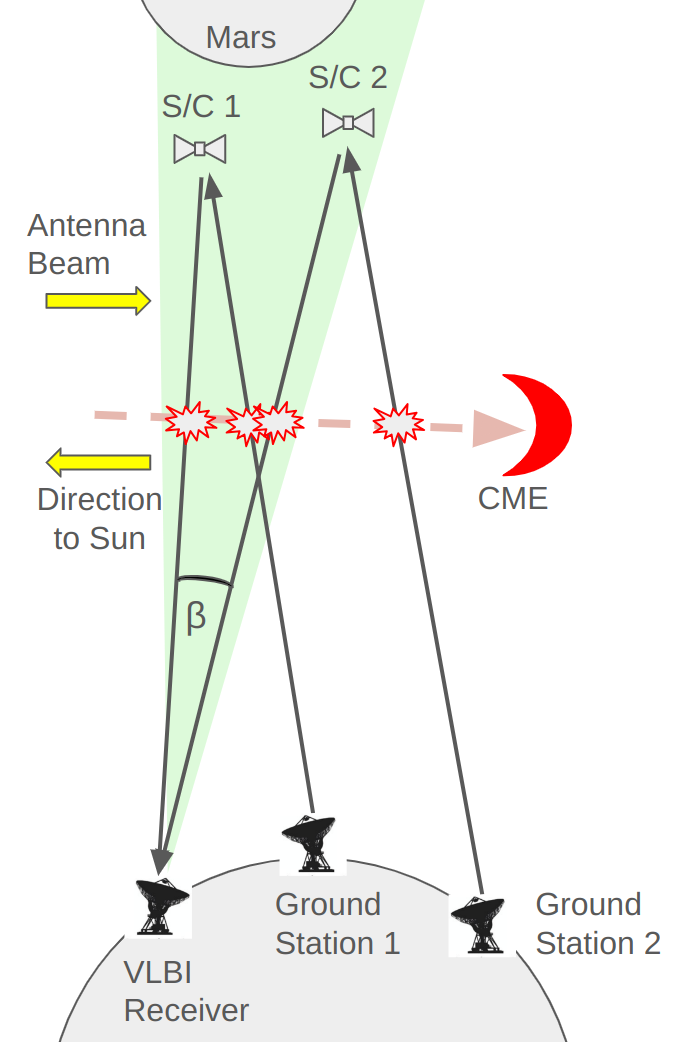}
\caption{Diagram of the multi-spacecraft-in-beam observing configuration. Note that the example spatial ordering of the spacecraft and radio antennas in the diagram is not necessarily reflective of ordering in the actual experiment.}
\label{multi_SC}
\end{figure}

Estimates of the bulk velocity of the large-scale plasma structures probed by phase fluctuations can be determined by measuring the time delay between perturbations \change{to}{of} the radio signal as the plasma inhomogeneities drift between adjacent radio ray-paths. Multiple radio ray-paths through the plasma can be achieved by using more than one receiving antenna or alternately, as is the case presented here, using multiple spacecraft. The location of VLBI antennas (downlink receivers), ground stations (uplink transmitters) and spacecraft are typically known to high degree of precision, meaning that the magnitude of the separation of the radio paths at an arbitrary point, including the primary scattering point, can be estimated accurately. For a single antenna with one primary beam, the angular separation of the spacecraft in the sky, $\beta$, must be less than the beam width of the receiving antenna to avoid nodding between target spacecraft. Next generation VLBI systems such as the VLBI Exploration of Radio Astrometry (VERA) system \citep{kobayashi2004vera} or phased array feed systems (PAFS) \citep{pereira2022next}, which are capable of forming multiple beams that can be steered electronically, make this requirement less strict. An example of observing geometry of the multi-spacecraft-in-beam method, as used in this experiment, is demonstrated in Figure \ref{multi_SC}. \add{Observations using the PRIDE technique are reliant on the spacecraft operating in two-way coherent communications configuration, where the spacecraft downlink signal is derived from the uplink signal on the ground, with reference to a high-precision clock. In this configuration, both the uplink and downlink carriers acquire
phase fluctuations when passing through solar plasma (as indicated by the red impact regions in Figure \ref{multi_SC}), the uplink phase fluctuations are transponded onto the downlink carrier and the observed phase fluctuations are cumulative of both uplink and downlink paths. The implications of this for the CCF measurements are not well understood.} \citet{woo1985doppler} \add{have suggested that the the received two-way fluctuations are simply the one-way fluctuations smoothed over a time scale of the round trip time. However, this may only effective if the time scale is small compared with that of the disturbance. It is also likely to be more effective if round trip time is large compared to the transit duration of the plasma inhomogeneities, which is not the case here.}

Determination of the drift time of the plasma structures is usually achieved through calculation of the normalised discrete linear cross-correlation function (CCF). For a finite time series of a real discrete parameter this is defined as:

\begin{equation}
\begin{aligned}\label{CCF_equation}
\textnormal{CCF} = \frac{1}{\sigma_{x}\sigma_{y}(N-1)} \sum_{m=0}^{N-1}x(m)y(m+k) \ ,\ \textnormal{CCF} \le 1
\end{aligned}
\end{equation}

where $\sigma_{x}$ and $\sigma_{y}$ are the standard deviations of the discrete time series $x(t)$ and $y(t)$ respectively. $N$ is the length, in samples, of the each of the time series being correlated and $k$ denotes the time offset, in samples, of time series $y(t)$. The autocorrelation function (ACF), $y(t) = x(t)$, is also useful for describing the variability of phase fluctuations and comparing how the structure of the plasma remains coherent.

For cross-correlation observations, the radio ray-path separations are restricted to distances shorter than the correlation distance of the observed scintillations. The spacecraft must also be sufficiently spatially separated that the transit of the scintillation pattern can be resolved, however, larger separations between the spacecraft can also result in lower correlation between the signals, particularly closer to the Sun. The correlation between inhomogeneities with small scales is less pronounced due to their shorter characteristic lifetimes and because the random component of the plasma velocity has a larger effect on the small-scale inhomogeneities than on those with larger length-scales. \citet{yakovlev1989occultation} observed that the Taylor hypothesis is highly appropriate for unperturbed solar wind plasma at distances of approximately $20\ R_{\odot}$. At this elongation, the maximum in the CCF was most apparent in the phase data and was not always at the same delay as the frequency data. Since frequency and phase data probe the large-scale fluctuations in the plasma density \citep{Coles1989}, the correlation distance for phase scintillations is at least $10^{6}$ km \citep{woo1977measuring}. More recently, \citet{cappuccio2024probing} observed that, at approximately $10\ R_{\odot}$, cross-correlations between spacecraft Doppler residuals (with different observing antennas) with lines-of-sight distance perpendicular to the solar radial axis up to the order of $4 \times 10^{5}$ km and a radial separation of $3.20 \pm 0.06 \ R_{\odot}$ still exhibited significant cross-correlations. However, the phase cross-correlation technique is not without its shortfalls, the main problem being that the ACF of phase is very wide, and so too is the CCF, meaning it is difficult to use the technique to measure a small time delay, corresponding to a comparatively small separation of the target spacecraft.

\subsection{Constraining the Location of the CME Impact Along the Radio Ray-path}

It is often assumed that the dominant perturbations to the radio signal are \change{imparted}{imprinted} at the point of closest approach of the radio ray-path to the Sun, hereafter referred to as the `pierce point'. For a nominal, spherically symmetric solar wind model, where the plasma density drops off rapidly with increasing distance from the Sun, this point has the highest plasma density along the radio-ray-path. It is also assumed that the radial velocity of the solar wind is perpendicular to the line-of-sight \add{at the pierce point}. However, for dense structures like CMEs, this is not necessarily the case.

When operating in two-way coherent mode, plasma inhomogeneities perturb both the uplink and the downlink of the radio signal. For a continuous time series with a duration longer than the round-trip light time of the radio signal (traveling to the spacecraft and back), the ACF can exhibit an `echo' peak at the time delay that corresponds to the time between the perturbation being caused on the downlink and the uplink \citep{Remus2014, armstrong1998radio, wohlmuth1997measurement}. The observed echo delay is given by:

\begin{equation}
\begin{aligned}\label{delay_correction}
T_{echo}  = \frac{2L}{c} \pm \frac{d}{v}
\end{aligned}
\end{equation}

where $L$ is the distance along the radio ray-path between the plasma induced perturbation and the spacecraft and $c$ is the speed of light. The term $\pm \frac{d}{v}$, where $d$ is the distance the impact point travels over the duration of the echo delay time and $v$ is the velocity of the plasma, is included as a first order correction for the drift of the primary scattering point caused by the movement of the Earth and the spacecraft. The correction is positive for the for the ingress period and negative for the egress period. Importantly, the echo delay may not always be present or apparent under all solar wind conditions and depends on the extent of the scattering region and the shorter characteristic lifetimes smaller scale fluctuations, particularly closer to the Sun \citep{wohlmuth1997measurement, richie2009space}. Critically, the determination of the the onset of the CME in the radio data, which is based on the initial and sudden rise in the scintillation parameters, is not affected by two-way data, because the uplink effect is not echoed until at many minutes later \cite{woo1978radial} and so the impact region of the CME can be determined relatively accurately.

Using the appropriate SPICE kernels \citep{acton2018look}, the positions of spacecraft and antennas, relative to other bodies such as the Sun, can be used to determine the direct radio paths between the antennas and spacecraft. The pierce point for the direct ray-path from each station can be determined by the vector projection of the spacecraft-Sun vector onto the spacecraft-VLBI antenna vector. The impact location of the CME along the radio ray-path, $L$, can then be established by comparing the observed echo delay time in the correlation functions and the corrective term to the echo delay time at the pierce point. The CME impact point along the radio ray-path is then calculated as a fraction of the light time from the spacecraft to the pierce point.

\subsection{Observations and Data Processing}\label{method}
The measurements presented here were part of a broader campaign observing the solar wind using the PRIDE technique during the 2023 Mars superior solar conjunction. The observations were conducted with the University of Tasmania's 30 metre radio telescope located in Ceduna, South Australia (VLBI station code `Cd'). The observing method was essentially identical to Paper I, the only difference was the introduction of continuous tracking of the target spacecraft instead of the static 19-minute pointings and 1-minute slewing windows used previously. Continuous tracking allows for more flexible data processing around transmission dropouts. It also keeps the signal to noise ratio (SNR) at a more stable level because the target is not drifting across the beam. 

Since the fluctuations due to the plasma are frequency dependent, ideally the spacecraft have radio frequency communications within the same frequency band (so the scintillation comparable) and close enough together that the observing configuration at a VLBI station is straightforward to implement but far enough apart that the signal from each spacecraft are easily distinguishable. Differentiating between spacecraft can be achieved by identifying the nominal transmission frequency, or, when the signal frequencies ar similar, by examining the signal Doppler. In this instance, the spacecraft X-band radio signals are separated in frequency, 8.420 GHz (MEX) and 8.430 GHz (TIW), such that each signal was prescribed a dedicated 8 MHz channel for digitisation. The heterodyned (down-converted) and digitised spacecraft signals were recorded in standard VLBI data interchange format (VDIF) common to VLBI stations. The raw VDIF data was processed with the SDtracker software suite \citep{MoleraCalves2021}, exactly as in previous campaigns\add{, separating the data into consecutive 20-minute segments for iterative precision Doppler compensation and phase extraction}.

\add{The outputs from the SDtracker software are the spacecraft signal SNR, Doppler noise ($\Delta f$), and residual phase ($\Delta\phi$). Post processing computation (numerical implementation of Equations \ref{eq:scintillation_index} and \ref{CCF_equation}) was conducted using the appropriate methods from the fundamental algorithms for scientific computing package in Python (SciPy)} \citep{Virtanen2020SCIPY}.

\section{Analysis and Results}\label{results and analysis}
\subsection{Context From Ancillary Coronagraph Data}

\begin{figure}[hbt!]
\centering
\includegraphics[width=1\linewidth]{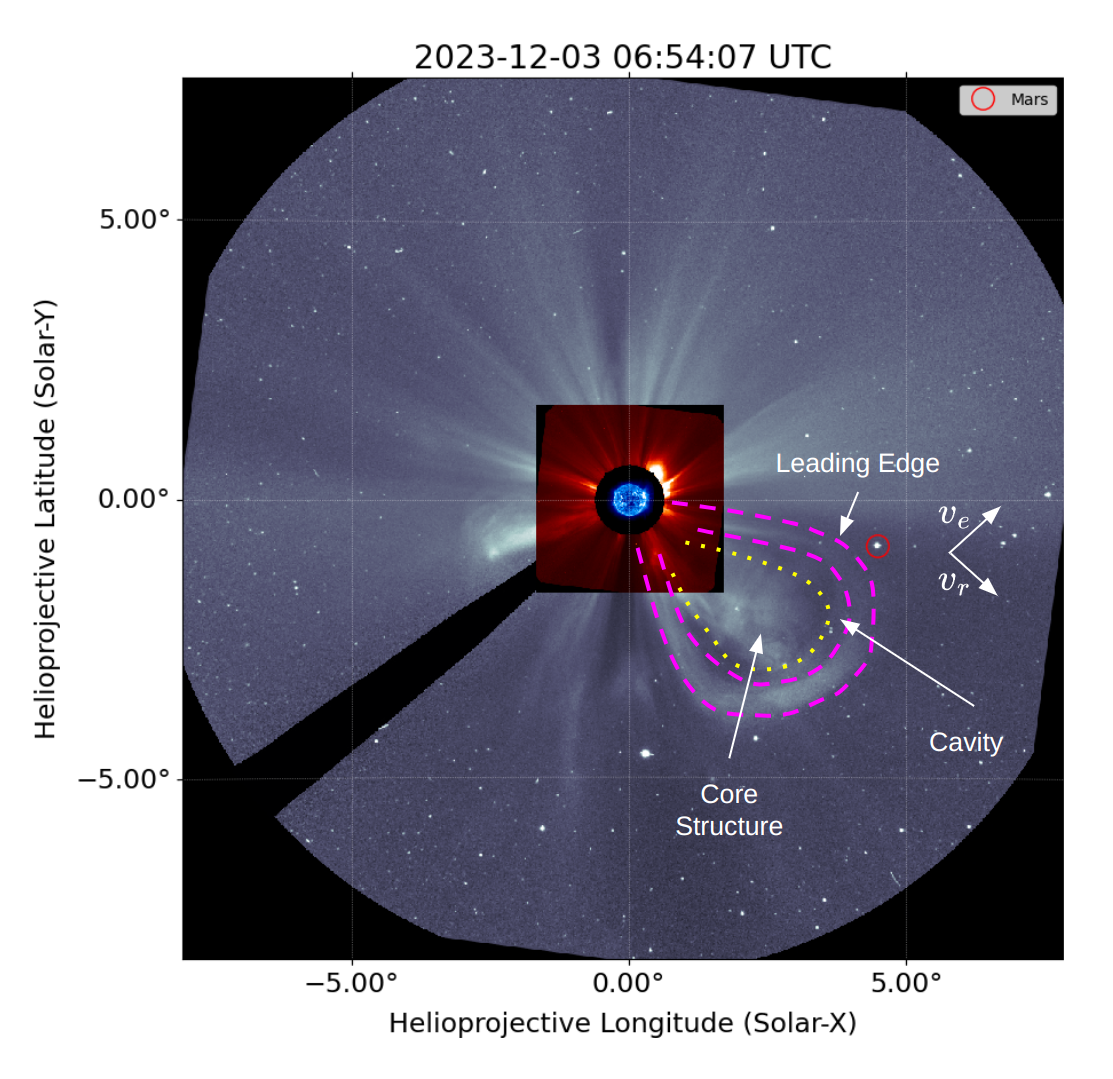}
\caption{A compound image of the CME event on 3$^{rd}$ December 2023 at 06:54:07. The image overlay displays capture from SECCHI EUVI (171 Angstroms) (centre), LASCO C2 (inner) and C3 (outer) coronagraphs. The classic three-part CME structure is annotated.}
\label{LASCO_C3}
\end{figure}

According to the SOHO LASCO CME Catalogue \citep{gopalswamy2024soho} and the M2M Catalogue \citep{evans2013score}, the CME first appeared in C2 imagery at 23:48 UTC on December 2$^{nd}$ and in C3 imagery at 01:42 UTC on December 3$^{rd}$. A snapshot of C2 and C3 imagery, captured roughly 8 minutes prior to the transit of the CME leading edge across the radio ray-path, is provided in Figure \ref{LASCO_C3}, including a centre overlay from the Extreme Ultra Violet Imager (EUVI) instrument package onboard STEREO Ahead spacecraft captured at 171 Angstroms. Each component of the classic three-part CME structure has been annotated. The highlighted borders of each feature are only rough guides and may not reflect the actual location of the CME structure -- especially where parts of the CME structure are only faint. The main purpose of the optical data presented here is to provide context to the radio science data.

\begin{figure}[hbt!]
\centering
\includegraphics[width=1\linewidth]{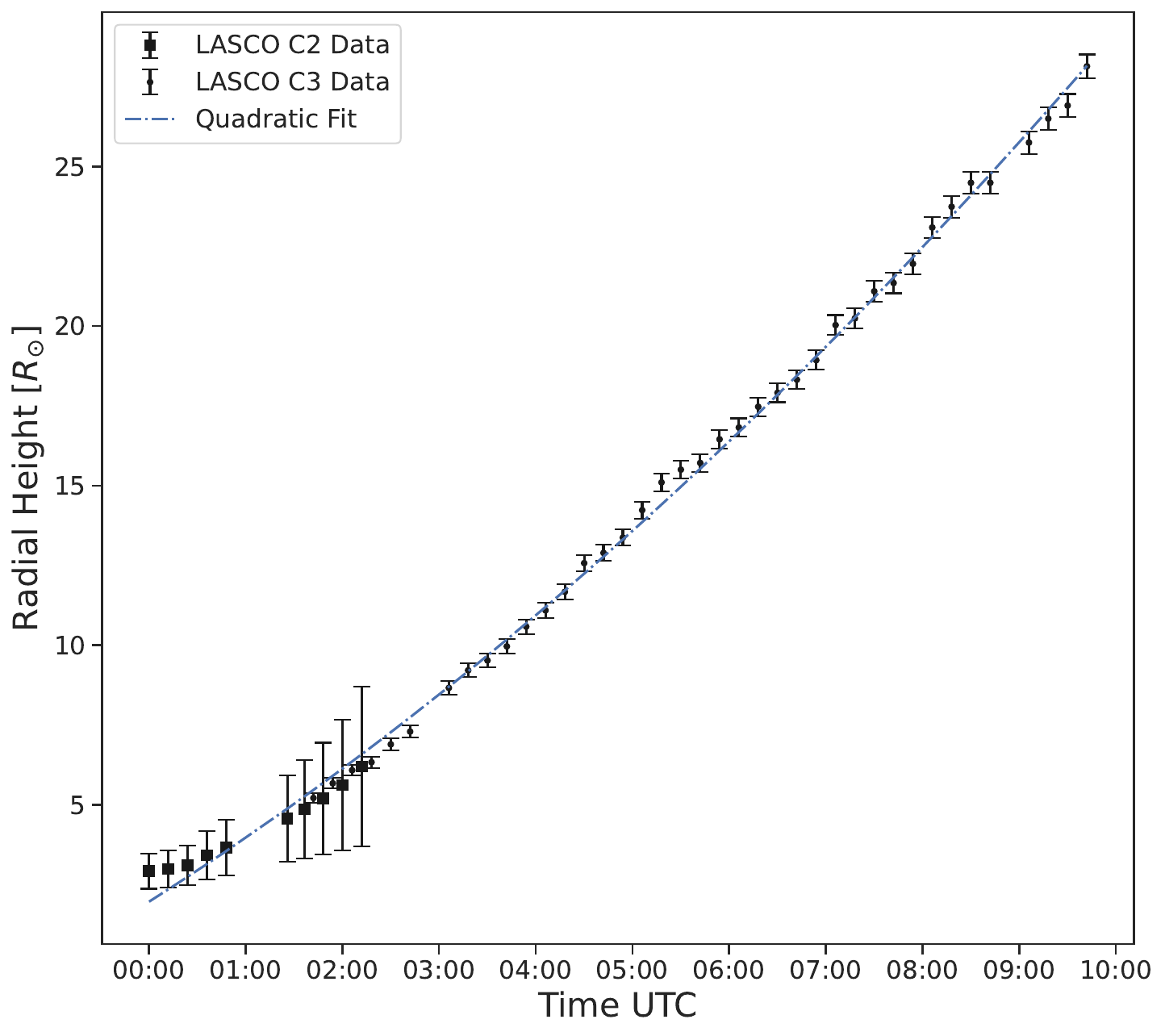}
\caption{LASCO C2 and C3 CME leading edge height-time plot (measurements from the SOHO LASCO CME Catalogue).}
\label{height_time}
\end{figure}

The CME originates from the Western solar limb with a central position angle (CPA) of $233^o$ degrees and an \add{initial angular width of $115^o$ (C2) and a revised angular width of $76^o$ (C3)}. Height-time data, obtained from the SOHO LASCO CME Catalogue and shown in Figure \ref{height_time}, tracks the furthest point of the leading edge from the Sun via running-difference imagery. Propagation of the \add{furthest point of the leading edge from the Sun is effectively radial (in direction $v_r$),} as viewed from the perspective of the LASCO instruments, so the CME edge reaches the equivalent radial offset of Mars before the leading edge front occults Mars. \add{The propagation direction of the CME plasma that occults Mars is not strictly radial (expanding in direction $v_e$) which may influence the shape and lag of the CCFs (\S \ref{SC_CCF}), although this is mostly dependent on the relative orientation and separation of the spacecraft in the plane of the sky.}

\add{The uncertainties of the height-time measurements in Figure \ref{height_time} were based on the models by} \citet{wen2007cme}, Equation \ref{LASCO_uncertainty}, \add{which take into consideration the spatial scale of pixels in the C2 and C3 instruments, the image cadence and the overall quality of the image (contrast of the leading edge structure relative to the background).}

\begin{equation}
\begin{aligned}\label{LASCO_uncertainty}
\frac{\delta(r)}{R_\odot} = \begin{cases}
0.065 \times \Bigl(\frac{\mathlarger{r}}{R_\odot}\Bigl)^2 & \text{, for C2}\\
\\
0.07 \times \sqrt{\Bigl(\frac{\mathlarger{r}}{R_\odot}\Bigl)} & \text{, for C3}\\
\end{cases}
\end{aligned}
\end{equation}

where $r$ is the distance of the radial height from the Sun.

The instantaneous gradient of a polynomial fit to the height-time data provides a measure of the propagation velocity of the CME leading edge. A linear fit (not-shown) yields an estimate of the CME leading edge expansion velocity as 516 km/s in the solar radial direction ($v_{r}$). A second order correction to this estimate with a quadratic polynomial (shown) yields a velocity at the final height within the coronagraph ($28\ R_{\odot}$) as 681 km/s and a velocity of 599 km/s at $20\ R_{\odot}$. Real-time analysis suggested that the CME eruption originated from an active region with an approximate heliospheric longitude of $130^o$ and latitude of $-34^o$, making it a far-side event. \change{This implies that the estimates of velocity from the coronagraph imagery and scintillation observations may underestimate the true CME velocity, since both techniques provide a weighted average of the plasma velocity perpendicular to the line-of-sight}{Consequently, estimates of the velocity derived directly from coronagraph imagery or scintillation techniques tend to underestimate the true CME velocity, since these techniques directly measure the projection of the true velocity onto the plane of the sky} \citep{alurkar1997solar}. Due consideration of the location of the CME origin at the solar `surface' and of echo delays in the two-way radio data can be used to correct this (see \S \ref{location}).

\subsection{Radio Science Data}
\subsubsection{SNR, Doppler Noise and Phase Residuals}
\begin{figure*}[hbt!]
\centering
\includegraphics[width=0.95\linewidth]{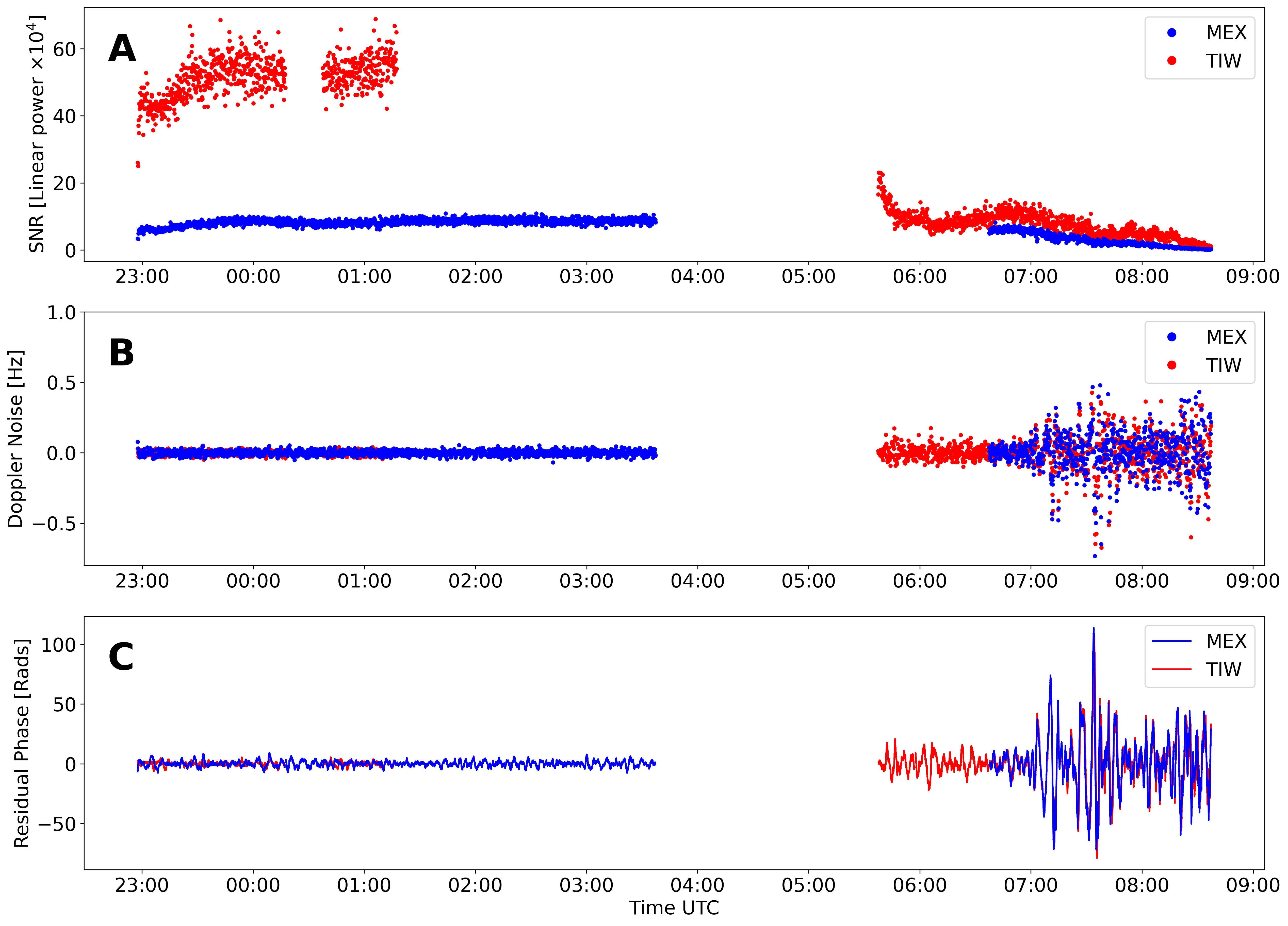}
\caption{SNR (A), residual frequency (10-second sample rate) (B) and residual phase (C) from MEX (blue) and TIW (red) spacecraft observed at Ceduna on December $3^{rd}$ 2023.}
\label{SNR_DOP_PHASE}
\end{figure*}

Substantial changes in SNR, Doppler noise and phase fluctuations are key indications of a CME across the spacecraft radio ray-path. Time series of these parameters for both MEX and TIW for the duration of the experiment are shown in Figure \ref{SNR_DOP_PHASE}. Tracking of both spacecraft began at 22:57 UTC on December 2$^{nd}$. Gaps in the radio data correspond to breaks in the two-way transmission from each craft. Tracking of both spacecraft concluded at 08:37 UTC corresponding to the receiving antenna reaching its lower elevation limit ($6^{o}$), just prior to Mars dropping below the horizon.

The initial radio data (22:57-03:37 UTC) from both spacecraft displayed typical levels of scintillation for the given solar elongation (mean phase RMS of 1.6 radians). When the transmission resumed (05:37 UTC), the Doppler noise and phase fluctuations for both spacecraft had increased marginally (phase RMS of 4.4 radians), before the passage of the CME leading edge at 07:02 UTC, with a peak phase RMS of 21.2 radians observed between 07:17-07:37 UTC. When the transmission from TIW resumed, the SNR was significantly lower than the initial radio data, likely due to manipulation of the output power onboard the spacecraft rather than plasma effects. The SNR of each spacecraft gradually dropped off from the onset of the CME until the conclusion of the experiment. The same features are evident in both the Doppler noise and the phase residuals. The impact of the CME leading edge on the radio data is quite clear and is consistent with the ancillary data from the LASCO C3 imagery. Further inferences about the structure and velocity of the plasma from a more detailed analysis of the radio data are provided in the following sections.

\subsubsection{Phase De-trend Comparison}
\begin{figure*}[h!]
\centering
\includegraphics[width=0.9\linewidth]{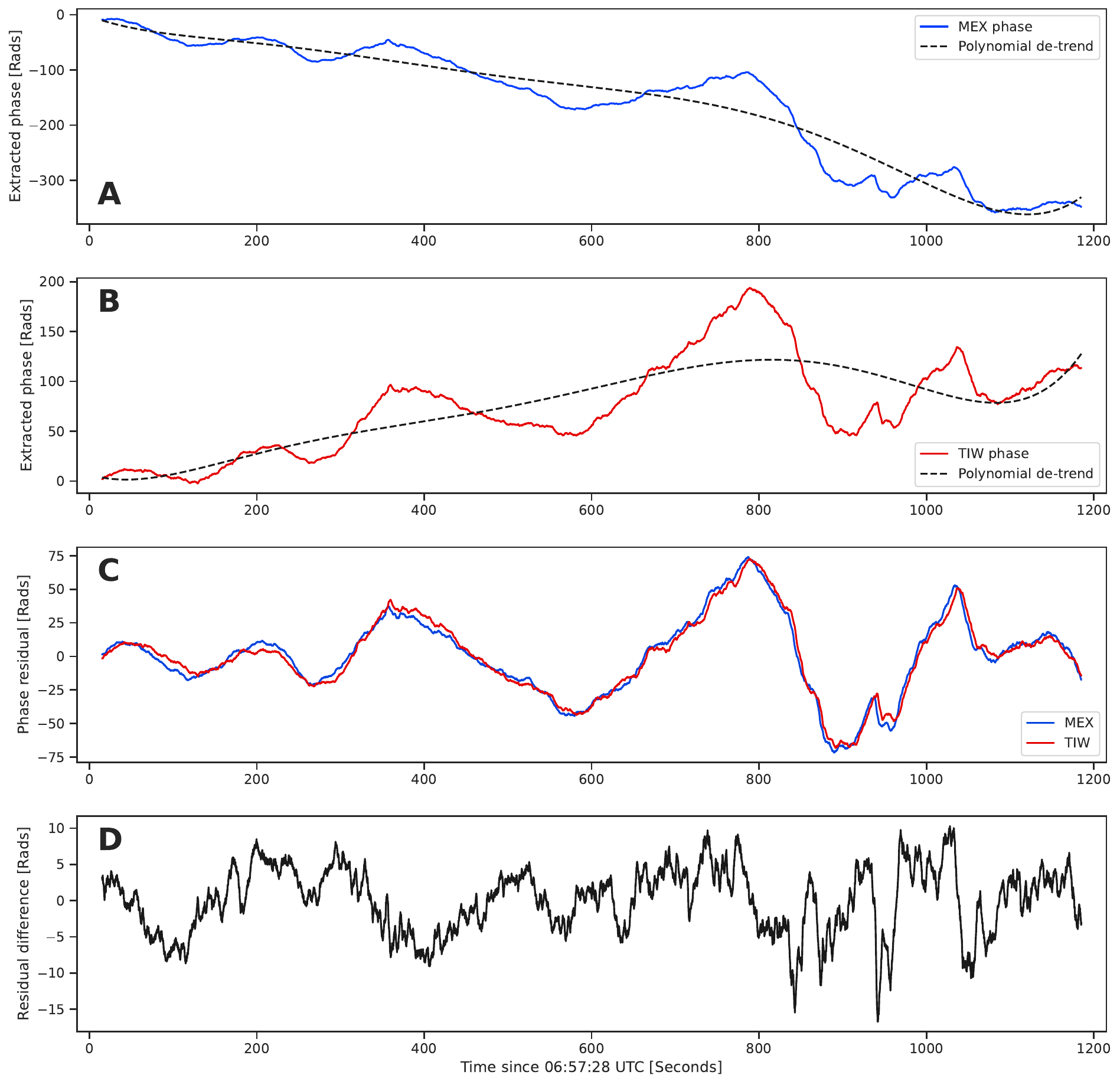}
\caption{Example extracted phases and associated polynomial fits for MEX (A) and TIW (B), residual phase \remove{for }for MEX and TIW after removal of respective polynomial fits (C) and the difference between MEX and TIW phase residuals (D).}
\label{phase_polyfit}
\end{figure*}

The extracted phases from MEX and TIW and their associated 6$^{th}$ order polynomial de-trends are shown in Figure \ref{phase_polyfit} A and B respectively. The two stage Doppler compensation ensures the spacecraft signal is static over a 20-minute duration \cite{MoleraCalves2021}. The order of the precision de-trend polynomial is dictated largely by the type and stage of the orbit, the proximity of the radio path to the Sun and the radio transponder ratio (as discussed in \citetalias{edwards2025studying}). A comparison of the Doppler compensation from both precise ephemerides and from polynomial fitting (as applied to MEX) can be found in \citet{ma2021vlbi}.

The interval presented in Figure \ref{phase_polyfit} contains the phase at the onset of the CME in the radio data at 07:02 UTC, where the CME perturbation of the downlink arrives first. Despite having different phase trends, owing primarily to the different orbital trajectories and propagation paths through the plasma, the residual (fit-measured) phases from each spacecraft, Figure \ref{phase_polyfit} C, are highly correlated. Each spacecraft has an independent and spatially separate ground station with their own clock references, so the system noise characteristics, excluding those of the VLBI receiving antenna, will not be shared between the two signals. The fact that the phase residuals are near identical suggests that the plasma disturbances from the CME are very large and do not evolve significantly over the duration it takes to transit from the line-of-sight from one spacecraft to another. The difference between phase residuals from each spacecraft are shown in Figure \ref{phase_polyfit} D. Shared noise characteristics between the phase residuals are suppressed by the difference and the time series retains a Kolmogorov power law index ($\alpha = 3.5 \pm 0.2$, see \citetalias{edwards2025studying} for a more extensive discussion of the PRIDE data and the power law index), indicating that the differences between the residual phases between the spacecraft is mainly owing to the different radio paths for each spacecraft and changes of the plasma turbulence at scales smaller than the separation of the radio paths.

\subsubsection{Multi-Spacecraft Cross-Correlation Analysis} \label{SC_CCF}

We preface the following discussion by clarifying that the ephemerides for TIW are not publicly available and so no calculations of the length or orientation of the projected baseline between radio ray-paths between the spacecraft was possible. Consequently, the full potential of the multi-spacecraft-in-beam technique has not been exploited here, however, the following analysis is still informative.

\begin{figure*}[hbt!]
\centering
\includegraphics[width=0.95\linewidth]{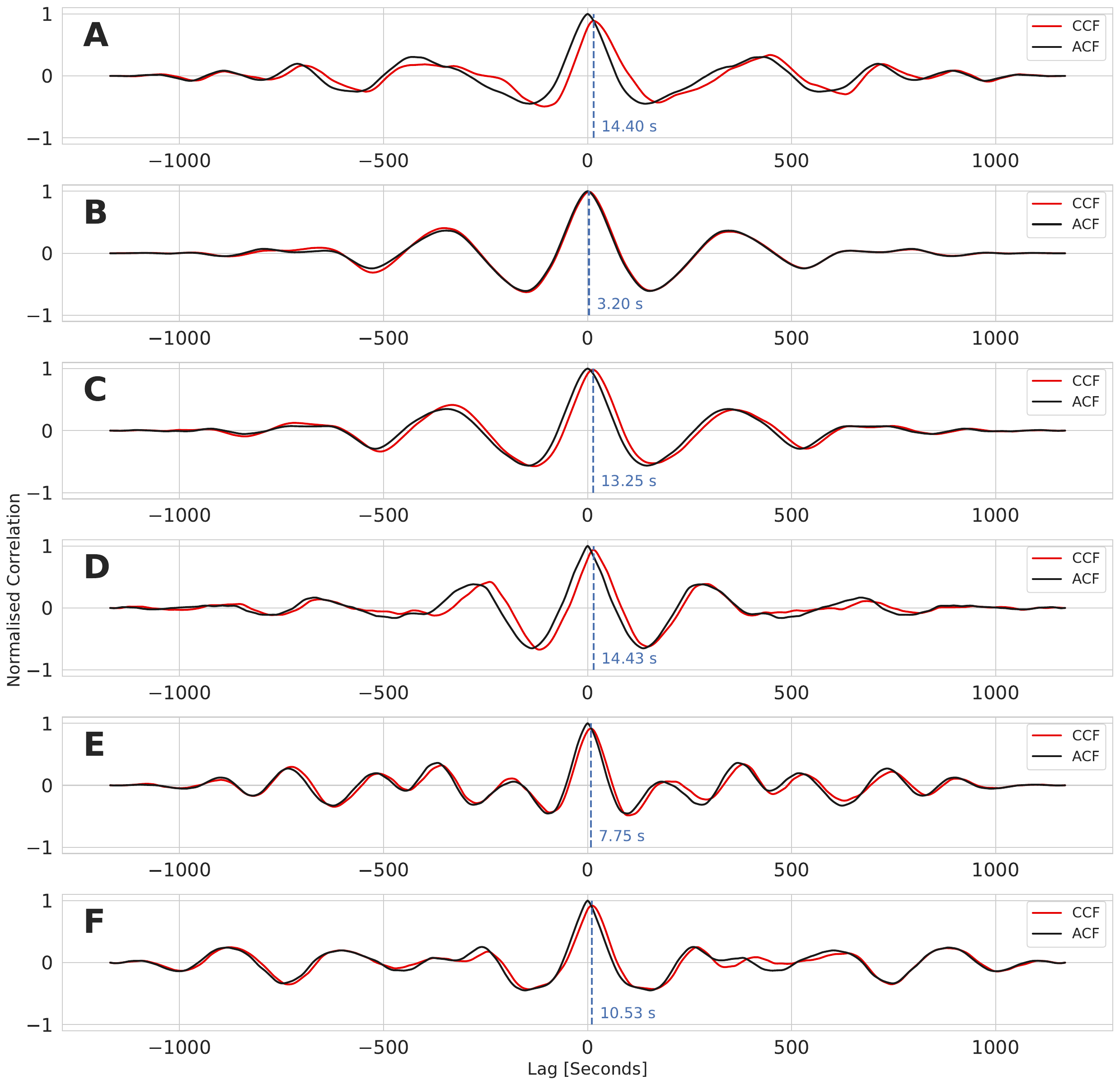}
\caption{CCF and ACF for consecutive 20-minute scans starting at 06:37 UTC (A) and concluding at 08:37 UTC (F). The vertical dashed line denotes the peak of the CCF.}
\label{CCF_ACF}
\end{figure*}

The ACF of the MEX phase (consecutive 20-minute intervals) and the CCF with the TIW phase (also 20-minute intervals) were calculated using Equation \ref{CCF_equation}. Upon inspection of the correlation functions several things become apparent. First, for the majority of overlapping intervals prior to 01:17 UTC, no meaningful \add{cross-}correlation was observed with the exception of two scans, one starting at at 23:57 UTC (correlation coefficient 0.84 and lag 6.05 s) and one starting at 00:57 UTC (correlation coefficient 0.64 and lag 3.18 s). The remaining correlation functions for 20-minute intervals from 06:37 UTC onwards are shown in Figure \ref{CCF_ACF}. The CCFs are almost identical (correlation coefficients $>0.89$) to the corresponding ACFs, just shifted in time. The magnitude of the correlation suggests that phase fluctuations are caused by exceptionally large plasma structures, especially compared to those at the beginning of the experiment, that do not evolve significantly as they drift between the separate radio ray-paths. The oscillations of the correlation functions about zero reflect fluctuations in the CME structure (density enhancements and diminishments) about the main increase in density across the CME leading edge. It is also possible that the orientation of the ray-path separation at this time was better aligned with the CME expansion velocity vector ($v_e$ in Figure \ref{LASCO_C3}) compared to velocity vector of the prevailing solar wind ($v_r$) prior to the CME onset, resulting in stronger correlation.

In all instances where the phases were correlated, the lags are biased in the same direction (positive), meaning that MEX was closer to the Sun than TIW over these periods. Interpreting the implications of the magnitude of the lag is more difficult since the general effect of the two-way nature of the radio signal on the delay of the correlation functions remains unclear. The uplink and downlink paths from each spacecraft have differing spatial separations and so the observed delay depends on whether perturbations \change{imparted}{imprinted} on the radio signal were more dominant on the uplink, the downlink or an aggregate of both. Where the fluctuations on the uplink and downlink are of comparable magnitude, the degree of correlation between the signals from each spacecraft may also be reduced. It is also likely that the breadth of the phase correlation peaks obscures any substructure, such as bimodality, caused by the different radio ray-path separations. Ultimately the effect on the CCF is highly dependent on the plasma structure and the observing geometry. 

Based upon the observed Doppler shift of each spacecraft signal and the similar magnitude of observed lags in panels A, C and D, we infer that the separation of the spacecraft does not change significantly over the duration of the experiment and the projected baseline remains relatively constant. The general lack of correlation prior to the CME onset was most likely caused by poor orientation of the separation of the radio ray-paths from each spacecraft relative to the prevailing solar wind direction, as was the case for \citet{tyler1981radio}. This means that only fluctuations larger than the scale of the ray-path separation will be correlated.

The onset of the CME is not obscured by the two-way effect and occurs in panel B with a lag of 3.2 seconds and a corresponding increase of the correlation coefficient (0.99) compared to earlier intervals. Based upon this, we infer that the smaller delays ($\approx$ 3.2 seconds) correspond to more significant perturbations \change{imparted}{imprinted} on the downlinks, the effects of which on the uplink will not arrive until $T_{echo}$ later. We propose that the larger delays ranging between 13.25-14.43 seconds correspond to the scenario where phases fluctuations on the uplink are larger than those on the downlink, so they are the dominant contribution to the ACF, and that the uplinks are more spatially separated. Consequently, the larger observed delays are caused by plasma drifting across more spatially separated radio uplink paths as opposed to corresponding to lower plasma velocities. Lags ranging from 6.05-10.53 seconds correspond to an aggregate of both the smaller and larger delays the demarcation of which is obscured by the breadth and smoothing effect of the CCF. Fluctuations of the plasma bulk velocity may also have contributed to the larger range of these intermediate lags. Knowledge of both the spacecraft positions is necessary for a more thorough understanding of the correlation and observed delays between the phase residuals from each spacecraft. \add{In the absence of orbital parameters for TIW, we defer more sophisticated analysis of the cross-correlation, including errors, to future work.} Additional research is also required to quantify the exact impact of the two-way signal on the correlation functions. However, the strong correlations observed here demonstrate that the multi-spacecraft-in-beam technique is a practical way of conducting multi-path propagation experiments when only one receiving antenna is available. 
 
\subsubsection{Comparison of Scintillation Index Levels with the Total Electron Content of the Background Solar Wind}\label{TEC_estimation}

\begin{figure}[hbt!]
\centering
\includegraphics[width=1\linewidth]{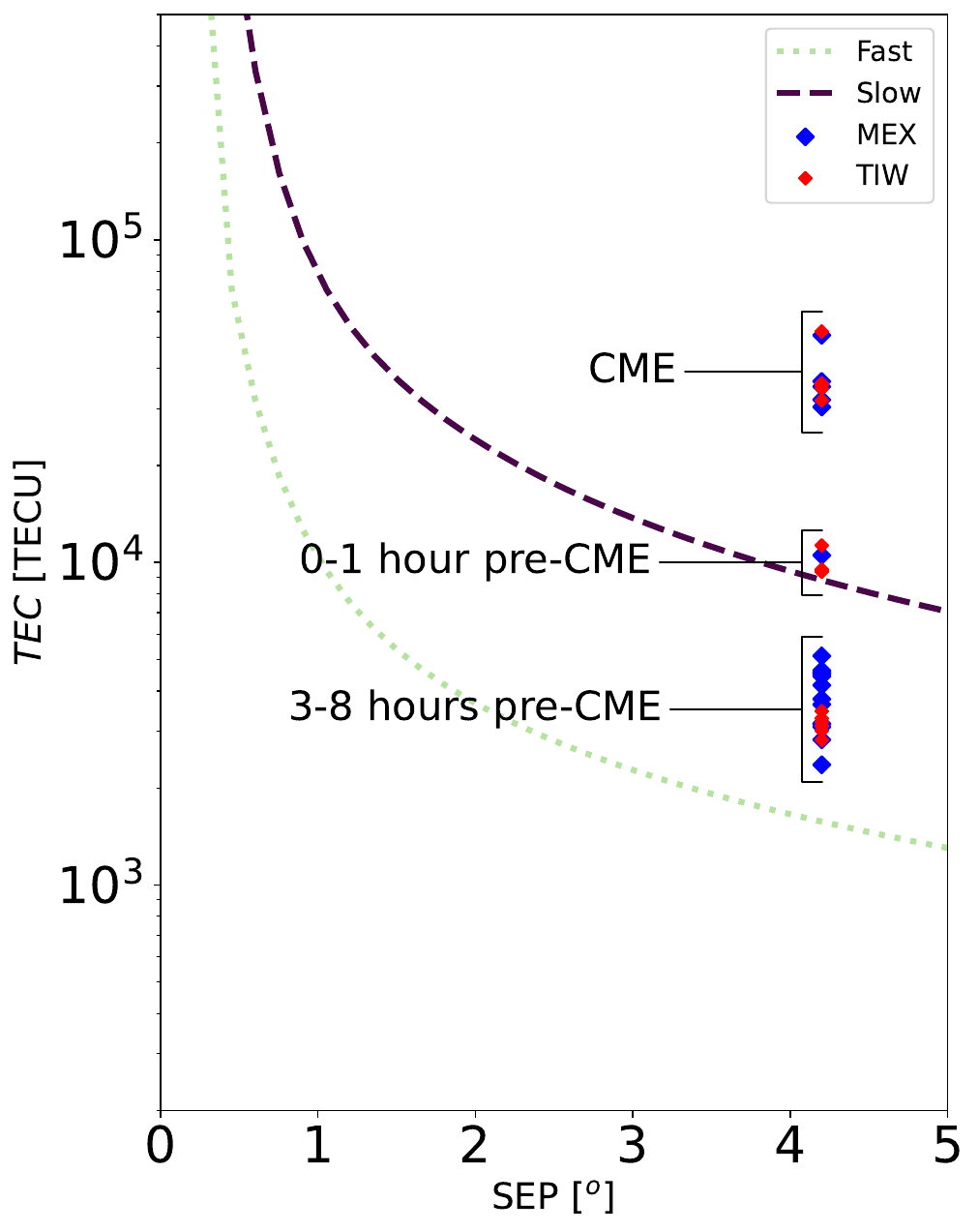}
\caption{The observed TEC from December $3^{rd}$ compared to Martian TEC model. TEC models for the fast (less dense) and slow (more dense) solar wind are denoted by the dotted and dashed lines respectively.}
\label{TEC}
\end{figure}

The TEC during each 20-minute interval was estimated using Equations \ref{eq:scintillation_index} and \ref{eq:TEC} and are shown in Figure \ref{TEC}. Comparison of the estimated TEC to the models, empirically derived for the background solar wind by \citet{Kummamuru2023}, shows three distinct levels of scintillation over the duration of the experiment. The first grouping occurs between 3-8 hours prior to the onset of the CME and displays values for the TEC  bounded by best fit models for the fast (less dense) solar wind and the slow (more dense) solar wind, typical for the solar elongation. The next grouping occurs for scintillation levels up to 1 hour prior to the onset of the CME. The estimated TEC was marginally elevated, approximately 2.4 times higher compared to the earlier conditions, and more closely match the slow solar wind model. This indicates some build up of electron content and turbulence prior to CME leading edge crossing the radio ray-paths. The final grouping corresponds to the onset and continued impacts of the CME leading edge where the estimated TEC levels were exceptionally higher than the models for the background solar wind. The observed scintillation levels during the CME were more comparable to levels observed at solar elongations more proximate to the Sun (within 1$^o$ SEP). While the linear relationship between $\sigma_{Sc}$ and TEC (Equation \ref{eq:TEC}) has been demonstrated for the solar wind, it is expected that the scaling relationship will break down in the case of a CME and should drastically overestimate the TEC, as was seen in \citet{Moleracalves2017}. The relationship between $\sigma_{Sc}$ and change in TEC is analysed in more detail in the following section.

The impact of a CME can be characterized by its strength, as described by the enhancement factor (EF) - the ratio of its peak to pre-transient or background scintillation level. With the onset of the CME, the scintillation jumps significantly, approximately 3.7 times higher than the scans immediately prior to the CME and 8.6 times higher than the initial conditions observed hours earlier. This is a relatively moderate value for the EF, which has been known to exceed 23 \citep{woo2007space}. It should be noted, however, that the radio signal is clearly probing a peripheral part or `flank' of the CME leading edge, not along the main propagation axis or the core structure, which would likely induce even more significantly elevated scintillation levels. The similarity in scintillation index values between the two spacecraft over simultaneous intervals demonstrates the technique is consistent in describing the plasma conditions, even when the phase fluctuations observed for each spacecraft are uncorrelated.

\subsubsection{Structure of the Leading Edge}
\begin{figure*}[hbt!]
\centering
\includegraphics[width=1\linewidth]{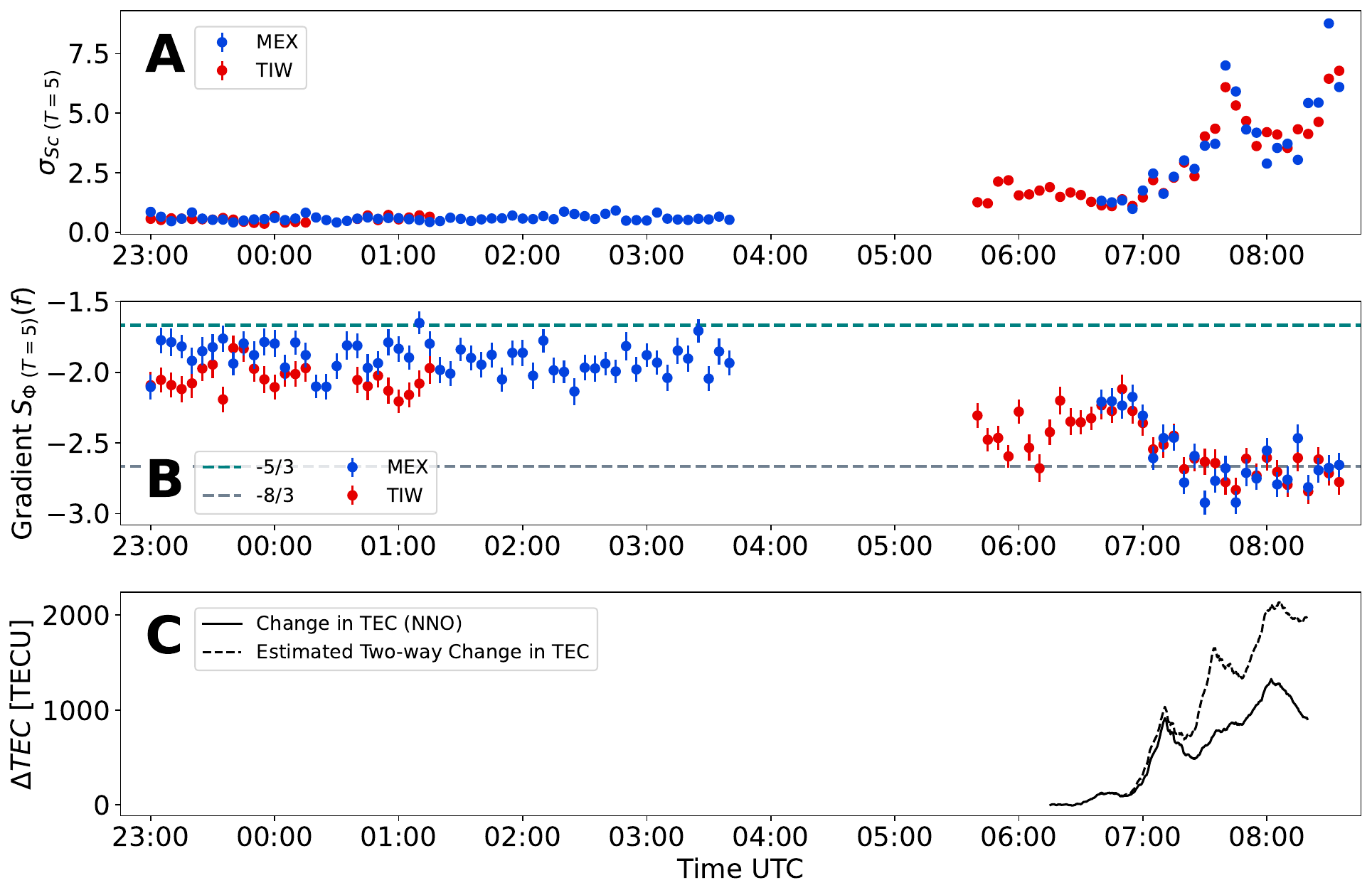}
\caption{$\sigma_{Sc}$ (A) and gradient of $S_{\Phi}(f)$ (B) calculated for 5-minute intervals. The middle plot has the theoretical slopes for 2D (-5/3) and 3D (-8/3) Kolmogorov turbulence for reference. Panel C displays the measured change in TEC derived from dual-frequency differential Doppler data from NNO (solid line) and an estimate of the two-way change in TEC (dashed line). }
\label{sigma_and_grad}
\end{figure*}

The temporal evolution of $\sigma_{Sc}$ provides insight into the structure of the CME leading edge. The change in $\sigma_{Sc}$ is due to a mixture of broadband changes of magnitude (vertical scaling), and changes in the slope of the phase power spectrum $S_{\phi}(f)$. The distribution of energy in the phase spectrum changes as the scattering plasma speed varies -- spectral energy shifts from low to high frequencies when the bulk velocity increases \citep{wheelon2001electromagnetic}. However, \citet{lynch2002comparison} demonstrated that the power spectrum profile can vary owing to the change in the apparent spatial scale due to the changing angle between the highly anisotropic structure and the plasma velocity vector, rather than just due to changes in bulk velocity. Figure \ref{sigma_and_grad} demonstrates the evolution of $\sigma_{Sc \ (T=5)}$ (A) and gradient of $S_{\Phi \ (T=5)}(f)$ (B) \add{over the duration of the experiment. The uncertainties for each parameter have were calculated using the methods established in \S \ref{background_scint}.}

The mean gradient in the period before the CME (23:00-03:45 UTC) \remove{(not shown) }was $-1.95 \pm 0.10$, closer to the Kolmogorov value for 2D turbulence (-5/3). Upon the resumption of the transmission from TIW (05:40 UTC), $S_{\Phi}(f)$ had decreased to below -2, roughly 20 minutes prior to the CME onset (07:02 UTC). There is a clear transition between 06:40 UTC and 07:30 UTC where the gradient of $S_{\Phi}(f)$ trends downwards stabilising around -8/3, the Kolmogorov value for 3D turbulence. This substantial change demonstrates that there was a clear difference turbulence regime of the plasma of the leading edge compared to the quiet period 3-8 hours prior.

The decrease in gradient of $S_{\Phi \ (T=5)}(f)$ affects the magnitude of $\sigma_{Sc \ (T=5)}$ to some extent, but the overall increase in $\sigma_{Sc \ (T=5)}$ is mainly due to a broadband increase in the strength of the phase fluctuations. This is associated with the non-stationary conditions at the arrival of the CME where there is a sudden dramatic changes in level of the turbulence structure constant \citep{Coles1989, wheelon2003electromagnetic}. An apparent spike in $\sigma_{Sc \ (T=5)}$ and a corresponding trough in gradient of $S_{\Phi}(f)$ occurred at 7:05. After the initial onset of the CME, there is a substantial increase in $\sigma_{Sc \ (T=5)}$ between 07:10-07:45 UTC. The power then decreases from 07:45, stabilising briefly at 08:00 UTC before increasing again at 08:15 UTC. $\sigma_{Sc \ (T=5)}$ is sensitive to the large changes in phase caused by substantial density fluctuations of the CME leading edge.

Simultaneously to the PRIDE experiment, albeit independently, transmission from MEX at X- and S-band (2.3 GHz) was also being recorded at the ESA's tracking station network (ESTRACK) New Norcia ground station (NNO) in Western Australia. The reduced closed-loop Doppler data from NNO, slightly different to the PRIDE data, is publicly available \citep{Paetzold2023odf} and supplement the PRIDE data in the following analysis. The frequency fluctuation measurements from each station are related and the large scale perturbations caused by the CME were apparent in both data sets. However, the lower temporal resolution of the frequency fluctuation data (1 second sample rate) relative to the short length of the projected baseline meant that precise drift delays could not be resolved, despite the favourable orientation of the baseline over the duration of the experiment. 

A major advantage of the dual-frequency Doppler data is the ability to determine the one-way change in TEC by isolating the dispersive (frequency dependent) effect of the plasma on the different downlink radio frequencies. The one-way change in TEC is shown in bottom panel of Figure \ref{sigma_and_grad}. The two-way change in TEC, as would have been encountered along the entire round-trip, has been estimated by summing a time-shifted copy of the one-way data by the echo delay identified in \S \ref{location} (1388 seconds).

The change in TEC began to rise relatively sharply at around 07:00 UTC corresponding to the onset of the CME. The structure of the leading edge revealed by the dual-frequency differential Doppler method is consistent with the general model proposed in \citet{patzold2012coronal} and correspond to large variations of increased electron content that precede the primary core structure of the CME. The spikes in $\sigma_{Sc \ (T=5)}$ occur lag behind the peaks in the $\Delta$ TEC because $\sigma_{Sc \ (T=5)}$ describes the plasma fluctuations over a time period and probes smaller scale fluctuations than the large scale trends that dominate the $\Delta$ TEC. This is due to the significantly higher sample rate (40 samples per second) of the phase fluctuations, the shorter duration of data used to estimate the parameter and the effects of the Doppler de-trend. Finally, it should be noted that the TEC estimates from \S \ref{TEC_estimation}, determined using the linear scaling relationship (Equation \ref{TEC}), overestimate the change in TEC of the CME (one-way change) by a factor of 20.

\subsection{Estimation of the CME Impact Location}\label{location}

The closed-loop Doppler data from NNO are frequency residuals, similar to the PRIDE Doppler noise parameter, and so the frequency fluctuations were used in the following analysis (instead of the phase residuals) to facilitate comparison between the measurements from each station. The round trip light time from the point of closest approach to the Sun (the solar wind pierce point) to MEX, estimated using geometry from the SPICE kernels, was 1512 seconds. The ACF of the NNO frequency residuals and the Cd Doppler noise were then calculated, using Equation \ref{CCF_equation}, for 40-minute intervals (2400 seconds). Example ACF of MEX two-way Doppler noise (frequency fluctuations) from Cd and NNO (both sampled at 1 second intervals) over two separate intervals -- one several hours prior to the CME (upper) and one starting just before the onset of the CME (lower) are shown in Figure \ref{NNO_Cd_CCF}. 

The shape of the ACF close to zero is important because without a peak close to zero, aside from the white-noise which has zero lag, there cannot be an echo of this feature at a higher lag. In the periods prior to the CME, the primary peak in the ACF occurs at zero lag and has no width (Figure \ref{NNO_Cd_CCF} top inlay) and is dominated by noise. Consequently, the lack of any apparent echo feature before the CME is mainly because the feature is obscured by noise. The absence of an apparent echo delay in normal solar wind conditions was also previously observed by \citet{wohlmuth1997measurement}.

In the period immediately after the CME onset, the ACF from both stations are markedly different, exhibiting greater width and structure of the primary peak, as well as at greater lags. The low frequency trends present in the NNO frequency residuals have been filtered out of the PRIDE Doppler noise by the high pass filter effect of the Doppler compensation process. This results in the deviations between the two ACFs and the marginally narrower correlation peaks in the Cd ACF but, at the finer scale, the ACF from Cd and NNO are very similar. A number of commensurable maxima occur in the ACF of the PRIDE data due to the fact that the Doppler noise parameter varies about zero, since it is a residual. A distinguishing feature of the echo delay is its stability, varying only with changes to the location of the primary scattering region - such as a CME, compared to the other correlation peaks, which are more dependent on the structure of the plasma density fluctuations. After the onset of the CME, the most significant correlation peak common to both ACFs and in proximity to the expected delay for the pierce point occurs at 1384 seconds (Figure \ref{NNO_Cd_CCF} bottom inlay). The weaker maximum correlation \change{value}{coefficient} (0.2) likely results from the erosion of fluctuations \change{imparted}{imprinted} upon the uplink as the signal traverses to the spacecraft and then back through the highly turbulent CME structure, as discussed in the previous section.

The uncertainty of the impact position $L$ is primarily due to two separate factors. The first is the size and asymmetry of the correlation peak, which are related to the size and underlying structure of the CME traversing the radio ray-path. From this observation alone, there is a spatial and temporal ambiguity which means there is not enough information to resolve, in detail, what part of the CME caused the asymmetrical shape of the correlation peak. Instead, we attempt to constrain the extent along the radio ray-path which the CME affected, based upon the positive correlation. To achieve this, the profile of the peak was approximated by a Gaussian least squares fit, as was done by \citet{wohlmuth1997measurement}, shown for the data here in the lower panel inlay of Figure \ref{NNO_Cd_CCF}. The mean and standard deviation of this Gaussian model were then used to define a 95\% confidence interval about the central maximum, describing the range of time over which there was positive correlation ($1388 \pm 46$ seconds). The range of this confidence interval was in relatively good agreement with similar S-band observations conducted by \citet{armstrong1998radio}, which exhibited a localised scattering feature in the solar wind at a much larger solar elongation ($139^o$) with a range of order 150 seconds. 

\begin{figure}[hbt!]
\centering
\includegraphics[width=1\linewidth]{figs/John_NNO_Cd_ACF_preANDduring.png}
\caption{Example normalised ACF for Cd Doppler noise (blue) and NNO X-band frequency residuals (black) from before the CME (top) and the period containing the CME onset (bottom). The round trip echo delay is not apparent in the period before the CME, however, a feature, inferred to be the echo delay, is present in the period containing the CME. The maxima of the Gaussian fit to the ACF of the Cd Doppler noise (red) occurs at 1388 seconds.}
\label{NNO_Cd_CCF}
\end{figure}

\begin{figure}[hbt!]
\centering
\includegraphics[width=1\linewidth]{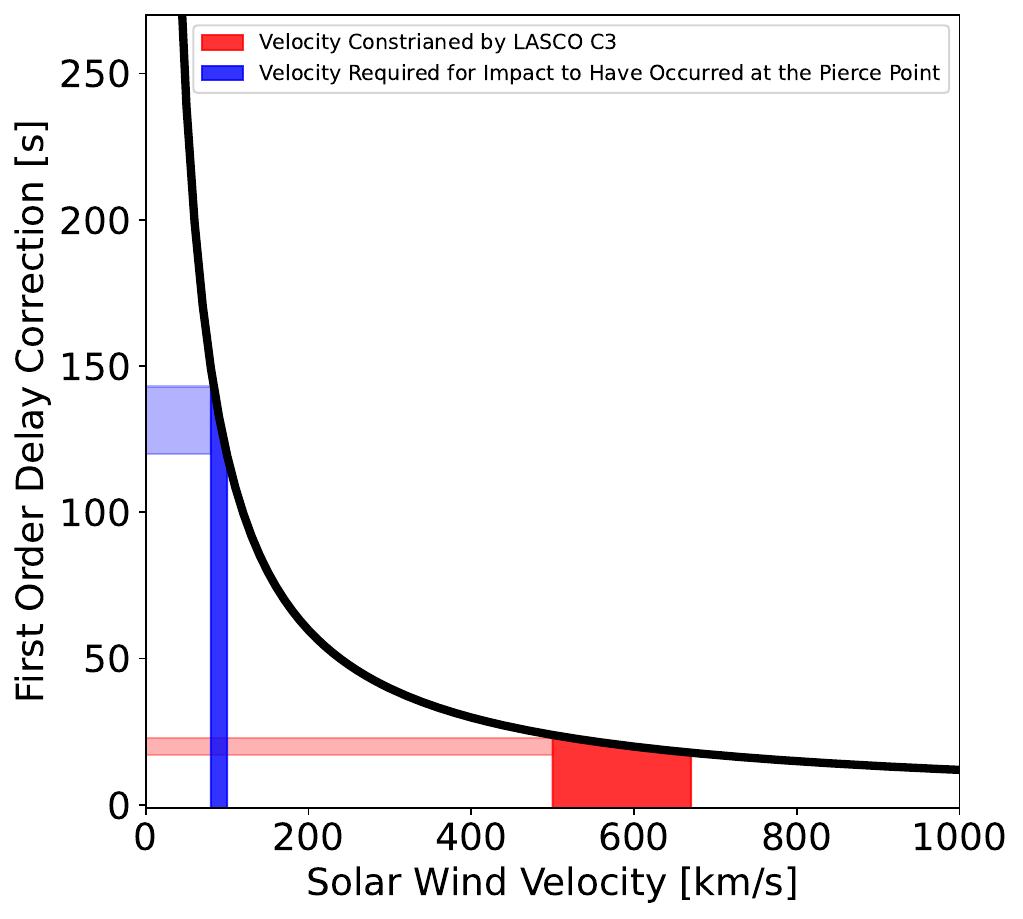}
\caption{First order correction delays for a range of solar wind velocities.}
\label{first_order}
\end{figure}

The second major uncertainty is the assumed solar wind velocity which is required to determine the first order correction term in Equation \ref{delay_correction}. For a far side CME event traversing the radio ray-path in the egress period, the correction term is added to the observed time delay, having the effect of moving the estimated location of the primary scattering location closer to the observer (Earth). The drift of the solar wind pierce point, due to the relative velocity of the Earth with respect to MEX, is 8.6 km/s. Over the duration of the estimated round trip echo delay (1388 seconds) this corresponds to a distance of approximately 12000 km. Without direct measurements, the solar wind velocity must then be assumed. The monotonic relationship between velocity and the first order delay correction over a wide range of typical velocities for the solar wind is provided in Figure \ref{first_order}. The measurements from LASCO C3 constrain the CME velocity in the plane of the sky to between 500-680 km/s, this results in a timing correction of between 20-24 seconds. For the perturbation to have occurred at the solar wind pierce point, the correction factor would require a solar wind velocity of approximately 97 km/s which is unrealistically slow at 20 $R_{\odot}$.

\begin{figure}[hbt!]
\centering
\includegraphics[width=0.95\linewidth]{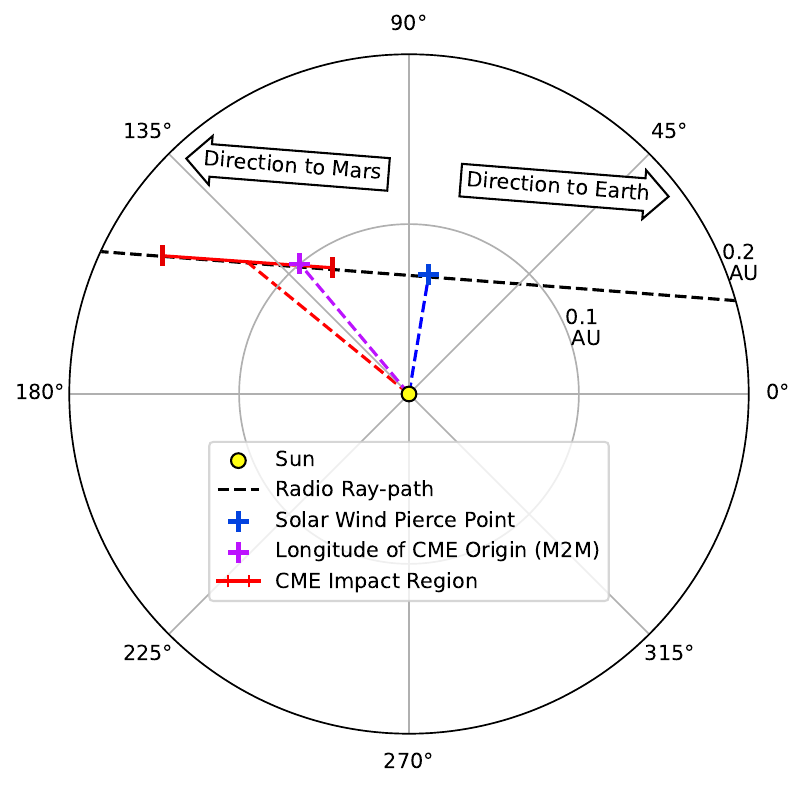}
\caption{Estimated location of the CME across the radio ray-path (red), based on the echo delay, as plotted in Heliographic-Stonyhurst coordinates. The radial spear from the origin identified in the M2M Catalogue (purple) and the solar wind pierce point (blue) at the closest approach are also marked.}
\label{Impact}
\end{figure}

A 10\% reduction in the first order correction, accounting for the shorter drift distance of the CME impact region compared to the solar wind pierce point ($\approx 1388/1512=0.92$), results in a first order correction factor of between 18-22 seconds. This results in an increase of the impact distance along the radio ray-path of the order of approximately 1.6\%. The region of scattering based on the echo delay, including the first order correction and the width of the correlation peak (46 seconds) is provided in Figure \ref{Impact}. The estimated extent along the radio ray-path that the CME transits is marked in red, along with a radial trajectory from the Sun. The solar wind pierce point at the closest approach to the Sun is marked in blue for reference. A radial propagation trajectory from the suspected point of origin of the CME at the solar surface, according to the M2M analysis, is marked in purple, the intercept of which with the radio ray-path is contained within the boundary of the estimated region of impact of the CME. The explicit cartesian coordinates of the centre of the CME impact region, lower bound (closer to Mars), and upper bound (closer to Earth) provided in AU in the cartesian reference frame native to the SPICE kernels are centre: (-0.11,-0.06,-0.03), lower bound: (-0.13,-0.10,-0.05) and upper bound: (0.09,-0.02.-0.01).

Given the launch time of the CME (23:48 UTC), the arrival time at the radio ray-path (07:02 UTC) and the distance traveled to reach the impact point on the radio ray-path (approximately 19648000 km) a revised velocity of the CME is at least $755 \pm 227$ km/s, where the uncertainty corresponds to the distance from the Sun to the upper and lower bounds of the impact region in Figure \ref{Impact}. The actual velocity is expected to be significantly higher because the radio signal was not probing the furthest point of the leading edge from the Sun. 

The ability to constrain the impact region of the CME along the radio ray-path of considerable value and is unique to the two-way spacecraft radio transmission observations. Providing additional trajectory and velocity information from real world events can be used to compare and improve CME propagation models and predictions. Conversely, ray tracing of MHD models may also help shed light on the impact of the CME structure on the two-way radio data and the shape of the ACF and help interpret the radio data. Future experiments should also seek to resolve the apparent absence of the echo peak in the case of the solar wind.

\section{Concluding Remarks}\label{conclusion}

Multi-path radio propagation experiments are a well established method for measuring coronal, solar wind and CME plasma. The work presented here demonstrates that the multi-spacecraft-in-beam technique, consisting of simultaneous radio transmission from multiple spacecraft and received at a single VLBI antenna, is an effective method for probing a CME leading edge structure. 

The frequency and phase residuals from both MEX and TIW spacecraft exhibited significant fluctuations caused by a CME leading edge across the radio ray-paths. Doppler compensation via iterative polynomial fitting was demonstrated to be highly effective method, accurately compensating for the effects of each spacecrafts orbital trajectory while retaining the plasma induced fluctuations in the signal frequency and phase. The phase residuals from each spacecraft were found to be highly correlated and exhibited clear delays associated with the passage of CME plasma material between the spatially separated radio ray-paths from each spacecraft. 

Estimates of the TEC derived from the phase scintillation index were shown to be consistent between each spacecraft for the for the background solar wind in the period prior to the CME. The scintillation index exhibited clear changes due to the passage of the CME leading edge but did not reflect the change in TEC very well. Changes in the slope of the phase power spectrum over the duration of the experiment indicated a clear change in the turbulence regime of the plasma. The primary cause of the increase in scintillation index was due to a broadband increase in the strength of fluctuation within the CME plasma. Dual-frequency differential Doppler data from NNO was used to establish the true change in TEC and provide a comparison and additional context to the PRIDE scintillation data. Ultimately the scintillation index remains a useful parameter for detecting an impact of a CME on the radio signal but more measurements are required to quantify the relationship between the scintillation index and the TEC in the case of a CME. \add{The use of alternative methods such as through comparisons with higher quality calibrated coronagraph images may be necessary to achieve this. However, part of the significance of the two-way radio propagation technique presented in this paper is the ability to determine the location of the CME along the line-of-sight. Determining electron density from white-light coronagraphs is complicated by the fact that the TEC derived from white-light intensity depends strongly on the assumed location of the electrons (see for example} \citet{vourlidas2006proper}). \add{The combination of white-light images with radio sounding data provide complementary information that can be used to improve estimates of TEC compared to white-light images alone.}\add{In addition, the wide variety of planned missions for monitoring solar activity, including the PROBA-3 mission, which features the ASPIICS coronagraph system with a field-of-view covering the very inner corona (1.1-3.0 $R_{\odot}$)}\citep{galy2023proba3}\add{; and the Polarimeter to Unify the Corona and Heliosphere (PUNCH) mission, which will include the Narrow Field Imager solar coronagraph to capture images of the outer corona at all position angles over solar elongations ranging from $1.5^o$ ($6\ R_\odot$) to $8^o$ ($32\ R_\odot$) in addition to three independent wide field heliospheric imagers, which will capture images of the entire inner solar system at solar elongations ranging from $3^o$ ($12\ R_\odot$) to $45^o$ ($180\ R_\odot$)} \citep{deforest2025polarimeter} \add{will provide unprecedented coverage of CME events, from their origins in the inner Corona to their propagation out into the extended heliosphere. It is anticipated that, in the future, the comprehensive coverage provided by these instruments, combined with radio propagation techniques, will mean that a variety of methods for constraining CME characteristics like TEC, acceleration and velocity, orientation and trajectory can be employed together to form a comprehensive and unified picture of CME events.}

With the extension of MEX's expected operation lifespan from 2025 to 2034, simultaneous observations of MEX and TIW for space weather studies can be continued. Future mission combinations where the multi-spacecraft-in-beam technique can be exploited include ESA's Mercury Planetary Orbiter (MPO) and JAXA's Mercury Magnetospheric Orbiter (Mio) spacecraft of the BepiColombo mission or ESA's JUICE and NASA's Europa Clipper spacecraft which will both be operational in the Jovian system with an overlapping window. The PRIDE technique is a formal component of the JUICE mission and large amounts of radio data will be collected over the missions lifespan making it a good candidate for space weather radio propagation studies. Importantly, the ephemerides for these missions are publicly available, so the exact geometry of the radio ray-paths can be determined and the full potential of the multi-spacecraft-in-beam technique can be exploited.

\begin{funding}
This research was supported by scholarship funding from the Commonwealth Scientific and Industrial Research Organisation of Australia (CSIRO) and from the Australian Government Research Training Program.
\end{funding}

\begin{acknowledgement}
The authors would like to thank Bev Bedson for tireless operation of the University of Tasmania Ceduna Radio Observatory; Rick Blake (European Space Agency) and Maoli Ma (Shanghai Astronomical Observatory) for sharing transmission window scheduling; and Bill Coles, Meng Jin (Lockheed Martin Solar \& Astrophysics Laboratory), Ron Ekers (CSIRO: Space and Astronomy) and Oliver White (UTAS) for constructive discussions regarding the experiment.

The core software suite (SDtracker version 2.0.14) used to process and analyse the raw VDIF data is publicly available at: https://gitlab.com/gofrito/sctracker and described in detail in \citet{MoleraCalves2021}. Additionally, this research used version 4.1.7 (DOI:10.5281/zenodo.7850372) of the SunPy open source software package (The SunPy Community et al., 2020).
\end{acknowledgement}


\bibliography{paperII}


\end{document}